\documentclass[prd,reprint,showpacs,showkeys]{revtex4-1}
\usepackage{amsfonts}
\usepackage{amssymb}
\usepackage{amsmath}
\usepackage{graphicx}
\usepackage[font={footnotesize,it}]{caption}

\pdfoutput=1

\begin{document}

\title{Einstein-Born-Infeld black holes with a scalar hair in
three-dimensions}
\author{S. Habib Mazharimousavi}
\email{habib.mazhari@emu.edu.tr}
\author{M. Halilsoy}
\email{mustafa.halilsoy@emu.edu.tr}
\affiliation{Department of Physics, Eastern Mediterranean University, Gazia\u{g}usa,
north Cyprus, Mersin 10, Turkey. }
\date{\today }

\begin{abstract}
We present black hole solutions in $2+1-$dimensional Einstein's theory of
gravity coupled with Born-Infeld nonlinear electrodynamic and a massless
self-interacting scalar field. The model has five free parameters: mass $M$,
cosmological constant $\ell $, electric $q$ and scalar $r_{0}$ charges and
Born-Infeld parameter $\beta $. To attain exact solution for such a highly
non-linear system we adjust, i.e. finely tune, the parameters of the theory
with the integration constants. In the limit $\beta \rightarrow 0$ we
recover the results of Einstein-Maxwell-Scalar theory, obtained before. The
self interacting potential admits finite minima apt for the vacuum
contribution. Hawking temperature of the model is investigated versus
properly tuned parameters. By employing this tuned-solution as basis, we
obtain also a dynamic solution which in the proper limit admits the known
solution in Einstein gravity coupled with self-interacting scalar field.
Finally we establish the equations of a general scalar-tensor field coupled
to nonlinear electrodynamics field\ in $2+1-$dimensions without searching
for exact solutions.\qquad
\end{abstract}

\maketitle

\section{Introduction}

Since the pioneering work of Banados, Teitelboim and Zanelli (BTZ) \cite{1,2}
the subject of $2+1-$dimensional black holes has attracted much attention
and remained deservedly ever a focus of interest due to many reasons.
Further to the pure BTZ black hole powered by a mass and a negative
cosmological constant the strategy has been to add new sources such as
electric / magnetic fields from Maxwell's theory \cite{3}, rotation \cite%
{3,4} and various fields \cite{5,6,7}. This remains the only possible
extension due to the absence of gravitational degree of freedom in the lower
dimension. In this situation scalar field coupling to gravity, minimal or
nonminimal with self-interacting potential is one such attempt that may come
into mind (See \cite{8} and references cited therein). The Brans-Dicke
experience in $3+1-$dimensions with a vast literature behind suggests that a
similarly rich structure can be established in the $2+1-$dimensions as well.
In this line Henneaux et al \ \cite{9} introduced Einstein's gravity
minimally and non-minimally coupled to a self interacting scalar field.
Einstein's gravity conformally and non-minimally coupled to a scalar field
was studied by Hasanpour et al in \cite{10} where they presented exact
solutions and their Gravity / CFT correspondences. Also, rotating hairy
black hole in $2+1-$dimensions was considered in \cite{11,12,13,14,15,16,17}
while charged hairy black hole was introduced in \cite{18}. Our purpose in
this study is to employ self-interacting scalar fields and establish new
hairy black holes in $2+1-$dimensions in analogy with the dilatonic case 
\cite{18}. In doing this, however, we replace also the linear Maxwell
electrodynamics with the nowadays fashionable non-linear electrodynamics
(NED). In particular, our choice of NED is the one considered originally by
Born and Infeld (BI) \cite{19,20,21,22,23,24} with the hope of eliminating
the electromagnetic singularities due to point charges. The elimination of
singularities in the electromagnetic field unfortunately doesn't imply the
removal of spacetime singularities in a theory of gravity-coupled NED.
Rather, the spacetime singularities may undergo significant revision in the
presence of NED to replace the linear Maxwell's theory. Herein, the
singularity at $r=0$ remains intact but becomes modified, both in powers of $%
\frac{1}{r}$ and also with the addition of a term such as $\ln r.$ Let us
add that there are special metrics hosting gravity-coupled NED which are
free of spacetime singularities \cite{25,26,27,28,29,30,31,32,33,34,35}.

Our $2+1-$dimensional model investigated in this paper consists of a
non-minimally coupled scalar field (with a potential) coupled to gravity and
NED field. We introduce such a model first, by deriving the field equations
and solving them. Recently such a model has been considered similar to ours
in which the linear Maxwell theory has been employed \cite{8}. Our task is
to extend the linear Maxwell Lagrangian to the NED Lagrangian of Born-Infeld
in $2+1-$dimensions \cite{36,37,38}. In particular limits our model recovers
the results obtained before. The self-interacting potential $U\left( \psi
\right) ,$ as a function of the scalar field $\psi $ happens in a particular
solution to be highly non-linear whereas the scalar field itself is
surprisingly simple: $\psi ^{2}\left( r\right) =\frac{r_{0}}{r+r_{0}},$ in
which $r_{0}$ is a constant such that\ $0\leq r_{0}<\infty .$ The scalar
field is bounded accordingly as $-1\leq \psi \left( r\right) \leq 1,$ and is
regular everywhere. As a matter of fact the constant $r_{0}$ is the
parameter that measures the scalar charge (i.e. the scalar hair) of the
black hole in such a model. Similar to the scalar field the static electric
field $E\left( r\right) $ also happens to be regular in our gravity-coupled
NED model in $2+1-$dimensions. The potential $U\left( \psi \right) $ is
plotted for chosen parameters which yields projection of a Mexican hat-type
picture where the reflection symmetry $U\left( \psi \right) =U\left( -\psi
\right) $ is manifest. The minima of the potential may be considered to
represent the vacuum energy of the underlying model field theory. To choose
one of the vacua we need to apply spontaneous symmetry breaking which lies
beyond our scope in this work. In the final section of the paper we consider
the general formalism for scalar-tensor field coupled with a NED Lagrangian.
Finding exact solution for this case is out of our scope here, but in a
future work we shall attempt a thorough investigation in this direction with
possible exact solutions in $2+1-$dimensions.

The paper is organized as follows. In Section II we introduce the field
equations, present a particular solution (with details in Appendix A),
investigate the limits and study some thermodynamical properties. In Section
III we obtain a dynamic solution from the solution found in Section II and
investigate its limits. In Section IV we apply the conformal transformation
from Jordan to Einstein frame and without solving the field equations we
find a general picture of the theory. The paper ends with Conclusion in
Section V.

\section{Field equations and the solutions}

We start with the action ($8\pi G=1=c$)%
\begin{multline}
S=\frac{1}{2}\int d^{3}x\sqrt{-g}\left[ R-8\partial ^{\mu }\psi \partial
_{\mu }\psi \right. - \\
\left. R\psi ^{2}-2V\left( \psi \right) +L\left( F\right) \right] 
\end{multline}%
in which $R$ is the Ricci scalar, $\psi $ is the scalar field which is
coupled nonminimally to the gravity, $V\left( \psi \right) $ is a
self-coupling potential of $\psi $ and $L\left( F\right) $ is the NED
Lagrangian with the Maxwell invariant $F=F_{\alpha \beta }F^{\alpha \beta }$
and electromagnetic $2-$form $\mathbf{F}=\frac{1}{2}F_{\mu \nu }dx^{\mu
}\wedge dx^{\nu }$. Let us add that this action differs from the one
considered in \cite{8} by a scale transformation in the scalar field and the
important fact that $L\left( F\right) $ here corresponds to an NED
Lagrangian rather than the Maxwell Lagrangian. Variation of the action with
respect to $g_{\mu \nu }$ implies 
\begin{equation}
G_{\mu }^{\nu }=\tau _{\mu }^{\nu }+T_{\mu }^{\nu }-V\left( \psi \right)
\delta _{\mu }^{\nu }
\end{equation}%
in which%
\begin{multline}
\tau _{\mu }^{\nu }=8\partial ^{\nu }\psi \partial _{\mu }\psi -4\partial
^{\lambda }\psi \partial _{\lambda }\psi \delta _{\mu }^{\nu }+ \\
\left( \delta _{\mu }^{\nu }\square -\nabla _{\mu }\nabla ^{\nu }+G_{\mu
}^{\nu }\right) \psi ^{2}
\end{multline}%
and 
\begin{equation}
T_{\mu }^{\nu }=\frac{1}{2}\left( L\delta _{\mu }^{\nu }-4F_{\mu \rho
}F^{\nu \rho }L_{F}\right) 
\end{equation}%
where $L_{F}=\frac{dL}{dF}$. Variation of the action with respect to $\psi $
and vector potential $A_{\mu }$ yields the scalar field equation%
\begin{equation}
\square \psi =\frac{1}{8}\left( R\psi +\frac{dV}{d\psi }\right) 
\end{equation}%
and nonlinear Maxwell's equation%
\begin{equation}
d\left( \mathbf{\tilde{F}}L_{F}\right) =0
\end{equation}%
respectively, in which $\mathbf{\tilde{F}}$ is the dual of $\mathbf{F}$. Our
line element is static circularly symmetric given by%
\begin{equation}
ds^{2}=-f\left( r\right) dt^{2}+\frac{1}{f\left( r\right) }%
dr^{2}+r^{2}d\theta ^{2}.
\end{equation}

The NED which we study is the well known BI theory. The BI-Lagrangian is
given by%
\begin{equation}
L\left( F\right) =\frac{4}{\beta ^{2}}\left( 1-\sqrt{1+\frac{\beta ^{2}F}{2}}%
\right) 
\end{equation}%
in which $\beta \geq 0$ is the BI parameter \cite{36} such that in the limit 
$\beta \rightarrow 0$ the Lagrangian reduces to the linear Maxwell Lagrangian%
\begin{equation}
\lim_{\beta \rightarrow 0}L\left( F\right) =-F
\end{equation}%
and in the limit $\beta \rightarrow \infty $ it vanishes so that the general
relativity (GR) limit is found. We also note that our electrodynamic
potential is only electric and due to that in the BI Lagrangian the term $%
G=F_{\mu \nu }\tilde{F}^{\mu \nu }$ is not present. The nonlinear Maxwell
equation admits a regular electric field of the form%
\begin{equation}
E\left( r\right) =\frac{q}{\sqrt{r^{2}+\beta ^{2}q^{2}}}
\end{equation}%
in which $q\geq 0$ is an integration constant related to the total charge of
the black hole. As we have shown in the Appendix A, the field equations
admit solution to the field equations as follows 
\begin{equation}
\psi ^{2}=\frac{1}{1+\frac{r}{r_{0}}}
\end{equation}%
in which $r_{0}\geq 0$ is a constant, and 
\begin{multline}
f\left( r\right) =\left[ -M+\frac{q^{2}}{1+\beta ^{2}}-2q^{2}\ln \left( 
\frac{r+\sqrt{q^{2}\beta ^{2}+r^{2}}}{2+q\beta }\right) \right] \times  \\
\left( 1+\frac{2r_{0}}{3r}\right) +r^{2}\left( \frac{1}{\ell ^{2}}+\frac{2}{%
\beta ^{2}}\right) + \\
\frac{2r^{2}r_{0}}{3q\beta ^{3}}\ln \left[ \frac{1}{r}\left( q\beta +\sqrt{%
q^{2}\beta ^{2}+r^{2}}\right) \right] - \\
\frac{2r}{\beta ^{2}}\left( \frac{r_{0}}{3r}+1\right) \sqrt{q^{2}\beta
^{2}+r^{2}}.
\end{multline}%
The self-coupled potential is given by%
\begin{equation}
V\left( \psi \right) =-\frac{1}{\ell ^{2}}+U\left( \psi \right) 
\end{equation}%
in which 
\begin{multline}
U\left( \psi \right) =\left( \frac{1}{\ell ^{2}}-\frac{M+2q^{2}\ln A}{%
3r_{0}^{2}}+\frac{q^{2}}{3r_{0}^{2}\left( 1+\beta ^{2}\right) }\right) \psi
^{6}+ \\
\frac{2r_{0}\left( \psi ^{6}-1\right) \ln B}{3q\beta ^{3}}+\frac{2\psi
^{2}\left( q^{2}\beta ^{2}\psi ^{4}+\left( 1-3\psi ^{4}\right)
r_{0}^{2}\right) }{3r_{0}\left( q\beta \psi ^{2}+\sqrt{\Delta }\right) \beta
^{2}}+ \\
+\frac{2\psi ^{8}\left( 2q^{2}\beta ^{2}+3\beta qr_{0}+2r_{0}^{2}\right) }{%
3r_{0}\beta ^{2}\left( q\beta \psi ^{2}+\sqrt{\Delta }\right) }+ \\
\frac{2\left( \beta q\left( 2\psi ^{2}+1\right) q+3r_{0}\psi ^{2}\right)
\psi ^{4}\sqrt{\Delta }}{3\beta ^{2}r_{0}\left( q\beta \psi ^{2}+\sqrt{%
\Delta }\right) },
\end{multline}%
with the abbreviations 
\begin{equation}
\Delta =\left( r_{0}^{2}+q^{2}\beta ^{2}\right) \psi ^{4}+r_{0}^{2}\left(
1-2\psi ^{2}\right) 
\end{equation}%
\begin{equation}
A=\frac{r_{0}\left( 1-\psi ^{2}\right) +\sqrt{\Delta }}{\psi ^{2}\left(
2+q\beta \right) }
\end{equation}%
and%
\begin{equation}
B=\frac{q\beta \psi ^{2}+\sqrt{\Delta }}{r_{0}\left( 1-\psi ^{2}\right) }.
\end{equation}%
We note that in terms of $r$ we have 
\begin{equation}
A=\frac{r+\sqrt{r^{2}+q^{2}\beta ^{2}}}{2+q\beta }
\end{equation}%
and%
\begin{equation}
B=\frac{q\beta +\sqrt{r^{2}+q^{2}\beta ^{2}}}{r}
\end{equation}%
which are independent from the scalar charge $r_{0}$. For $q\beta
\rightarrow 0$ one obtains $A\rightarrow r$ and $B\rightarrow 1.$ Similarly%
\begin{equation}
\sqrt{\Delta }=\frac{r_{0}}{r+r_{0}}\sqrt{r^{2}+q^{2}\beta ^{2}}
\end{equation}%
which vanishes in the limit $r_{0}\rightarrow 0.$ The general solution found
here is a singular black hole solution whose limits and horizons will be
investigated in the rest of the paper.

\subsection{The Limits}

The solution given in (11)-(14) for different limits represents the known
solutions in $2+1-$dimensions. The first limit is given with $%
r_{0}\rightarrow 0$ which implies $\psi \rightarrow 0.$ In this setting one
finds%
\begin{multline}
f_{BI}\left( r\right) =\lim_{r_{0}\rightarrow 0}f\left( r\right) = \\
-M+\frac{q^{2}}{1+\beta ^{2}}-2q^{2}\ln \left( \frac{r+\sqrt{q^{2}\beta
^{2}+r^{2}}}{2+q\beta }\right)  \\
+r^{2}\left( \frac{1}{\ell ^{2}}+\frac{2}{\beta ^{2}}\right) -\frac{2r}{%
\beta ^{2}}\sqrt{q^{2}\beta ^{2}+r^{2}}.
\end{multline}%
which is the black hole solution in Einstein-Born-Infeld (EBI) theory
introduced by Cataldo and Garcia (CG) in \cite{36}. We notice that the
integration constants in the general solution are finely tuned such that in
the EBI limit the solution admits both BTZ and CG-BTZ limits without need
for a redefinition of the electric charge. Otherwise it can be seen in the
Eq. (29) of Ref. \cite{36} that the CG-BTZ limit has different charge from
the original solution (2) of \cite{36}. This form of the solution easily
gives BTZ and charged BTZ black holes in the limits when $\beta \rightarrow
\infty $ and $\beta \rightarrow 0$ respectively i.e.,%
\begin{equation}
f_{BTZ}\left( r\right) =\lim_{\beta \rightarrow \infty }f_{BI}\left(
r\right) =-M+\frac{r^{2}}{\ell ^{2}}
\end{equation}%
and%
\begin{equation}
f_{CG-BTZ}\left( r\right) =\lim_{\beta \rightarrow 0}f_{BI}\left( r\right)
=-M+\frac{r^{2}}{\ell ^{2}}-2q^{2}\ln r.
\end{equation}%
We note that the limit of the self-coupling potential when $\psi \rightarrow
0$ becomes%
\begin{equation}
\lim_{\substack{ \psi \rightarrow 0 \\ r_{0}\rightarrow 0}}V\left( \psi
\right) =-\frac{1}{\ell ^{2}}
\end{equation}%
which is nothing but the cosmological constant in the action.

The other limit of the solution is given by $\beta \rightarrow 0$ which
yields%
\begin{multline}
f_{XZ}\left( r\right) =\lim_{\beta \rightarrow 0}f\left( r\right) = \\
-M+\frac{r^{2}}{\ell ^{2}}-2q^{2}\left( 1+\frac{2r_{0}}{3r}\right) \ln r+%
\frac{2r_{0}}{3r}\left( \frac{q^{2}}{3}-M\right) 
\end{multline}%
which is the black hole solution in Einstein-Maxwell coupled scalar field
found by Xu and Zhao (XZ) in \cite{8}. This limiting solution in the further
limit $q\rightarrow 0$ becomes%
\begin{equation}
\lim_{q\rightarrow 0}f_{XZ}\left( r\right) =-M+\frac{r^{2}}{\ell ^{2}}-\frac{%
2r_{0}M}{3r}
\end{equation}%
and when $r_{0}\rightarrow 0$ gives%
\begin{equation}
\lim_{r_{0}\rightarrow 0}f_{XZ}\left( r\right) =-M+\frac{r^{2}}{\ell ^{2}}%
-2q^{2}\ln r
\end{equation}%
which is the charged BTZ in its original form. To complete our discussion we
also give the limit of the potential when $\beta \rightarrow 0$. This can be
found as%
\begin{multline}
\lim_{\beta \rightarrow 0}V\left( \psi \right) =-\frac{1}{\ell ^{2}}+\left( 
\frac{1}{\ell ^{2}}-\frac{M}{3r_{0}^{2}}\right) \psi ^{6}- \\
\frac{2q^{2}\psi ^{6}}{3r_{0}^{2}}\ln \left( \frac{r_{0}\left( 1-\psi
^{2}\right) }{\psi ^{2}}\right) - \\
\frac{q^{2}\psi ^{6}\left( 2\psi ^{4}+2\psi ^{2}-7\right) }{9r_{0}^{2}\left(
1-\psi ^{2}\right) ^{2}}.
\end{multline}%
We must add that due to the modification made in the action (1) our results
are much simpler than those given in \cite{8} but still with a redefinition
of the parameters and by rescaling the scalar field one recovers the forms
found in \cite{8}. To complete this section we give the form of Ricci scalar
in terms of the new parameters:%
\begin{equation}
R=\Pi r_{0}+\Xi 
\end{equation}%
in which%
\begin{multline}
\Pi =\frac{4\left( 2q\beta \left( q\beta +r\right) +r^{2}\right) }{\left(
q\beta +r\right) ^{2}\beta ^{2}\chi }-\frac{4\left( r+2q\beta \right) \chi
\varpi }{\beta ^{3}q\left( q\beta +\chi \right) ^{2}}- \\
\frac{4q\left( 2\chi +r\right) \varpi }{\beta \left( r+\chi \right) \left(
q\beta +\chi \right) ^{2}}+\frac{8\beta q^{3}}{B\left( r+\chi \right) \left(
q\beta +\chi \right) ^{2}},
\end{multline}%
and%
\begin{multline}
\Xi =\frac{4q^{2}\left[ 4q^{3}\beta ^{3}-2\left( \chi -2r\right) rq\beta
+\left( r^{2}+2q^{2}\beta ^{2}\right) \left( 2\chi -r\right) \right] }{\chi
\left( r+\chi \right) r\left( q\beta +\chi \right) ^{2}}- \\
\frac{6\left( 2q\beta \left( q\beta +r\right) +r^{2}\right) }{\left( q\beta
+\chi \right) ^{2}\ell ^{2}}-\frac{12q^{3}\beta ^{3}}{\left( r+\chi \right)
\left( q\beta +\chi \right) ^{2}\ell ^{2}}
\end{multline}%
in which $\chi =\sqrt{q^{2}\beta ^{2}+r^{2}}$ and $\varpi =\ln \left( \frac{%
q\beta +\sqrt{q^{2}\beta ^{2}+r^{2}}}{r}\right) .$ One easily observes that
for $\lim_{r_{0}\rightarrow 0}R=\Xi ,$ but to see the structure of the
singularity we expand $R$ about $r=0$ which gives%
\begin{multline}
R=\frac{8q}{\beta r}+\frac{4r_{0}}{q\beta ^{3}}\ln r+ \\
\frac{4r_{0}}{q\beta ^{3}}\left( 1-\ln \left( 2q\beta \right) \right)
-6\left( \frac{2}{\beta ^{2}}+\frac{1}{\ell ^{2}}\right) +\mathcal{O}\left(
r\right) .
\end{multline}%
This shows that the singularity is of the order $\frac{1}{r}$ which, apart
from the logarithmic term is weaker than the Einstein-Maxwell-Scalar
solution \cite{8}, which was of the order $\frac{1}{r^{3}}$.

\subsection{Horizon(s) and Hawking temperature}

\begin{figure}[h]
\includegraphics[width=65mm,scale=0.7]{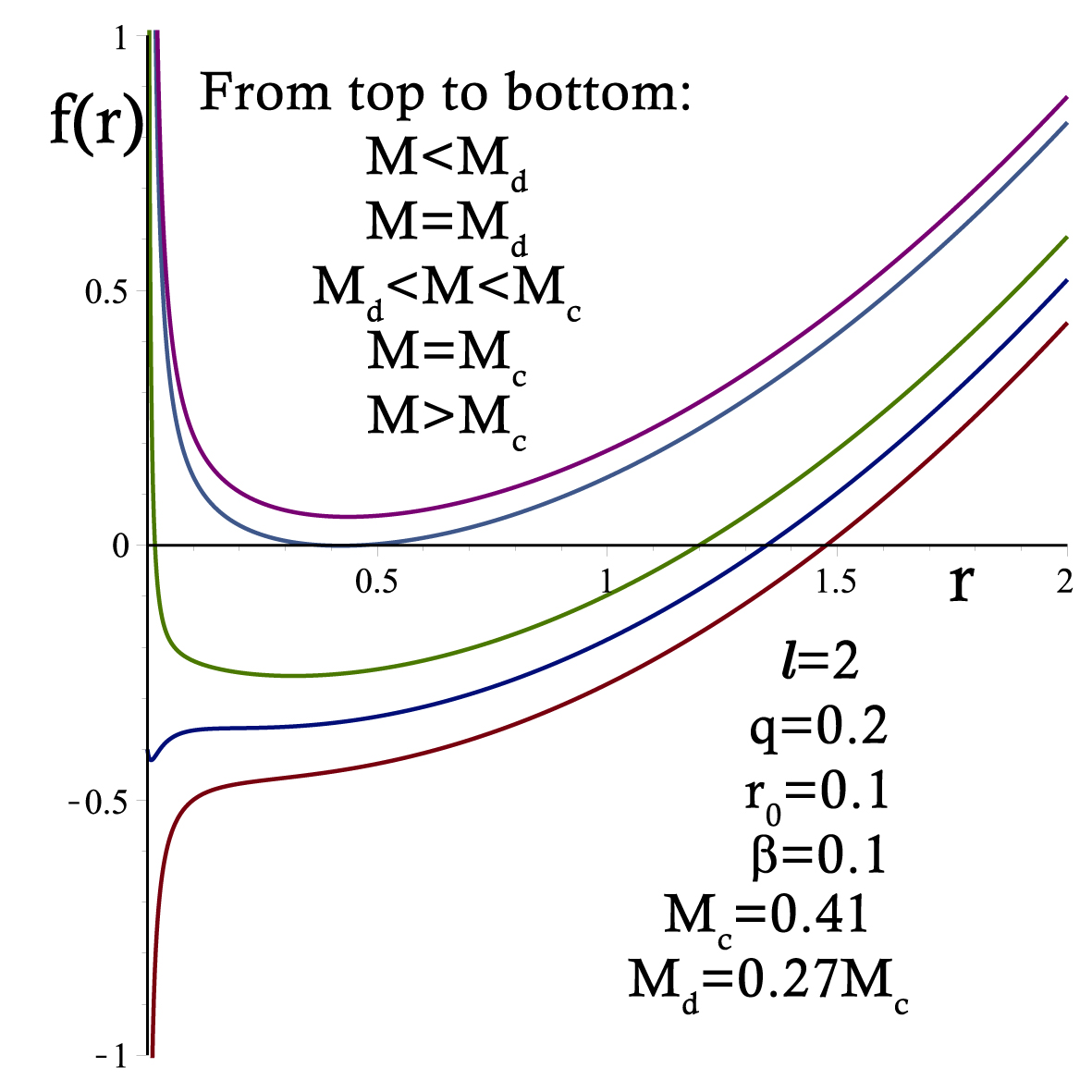}
\caption{Metric $f\left( r\right) $ function versus $r$ in terms of
different masses for the black hole. As it is clear the mass of the central
object must be bigger than a certain mass to have black hole solution. Note
that this is valid for non-zero cosmological constant i.e., $\frac{1}{\ell
^{2}}\neq 0.$ }
\label{fig: 1}
\end{figure}
\begin{figure}[h]
\includegraphics[width=65mm,scale=0.7]{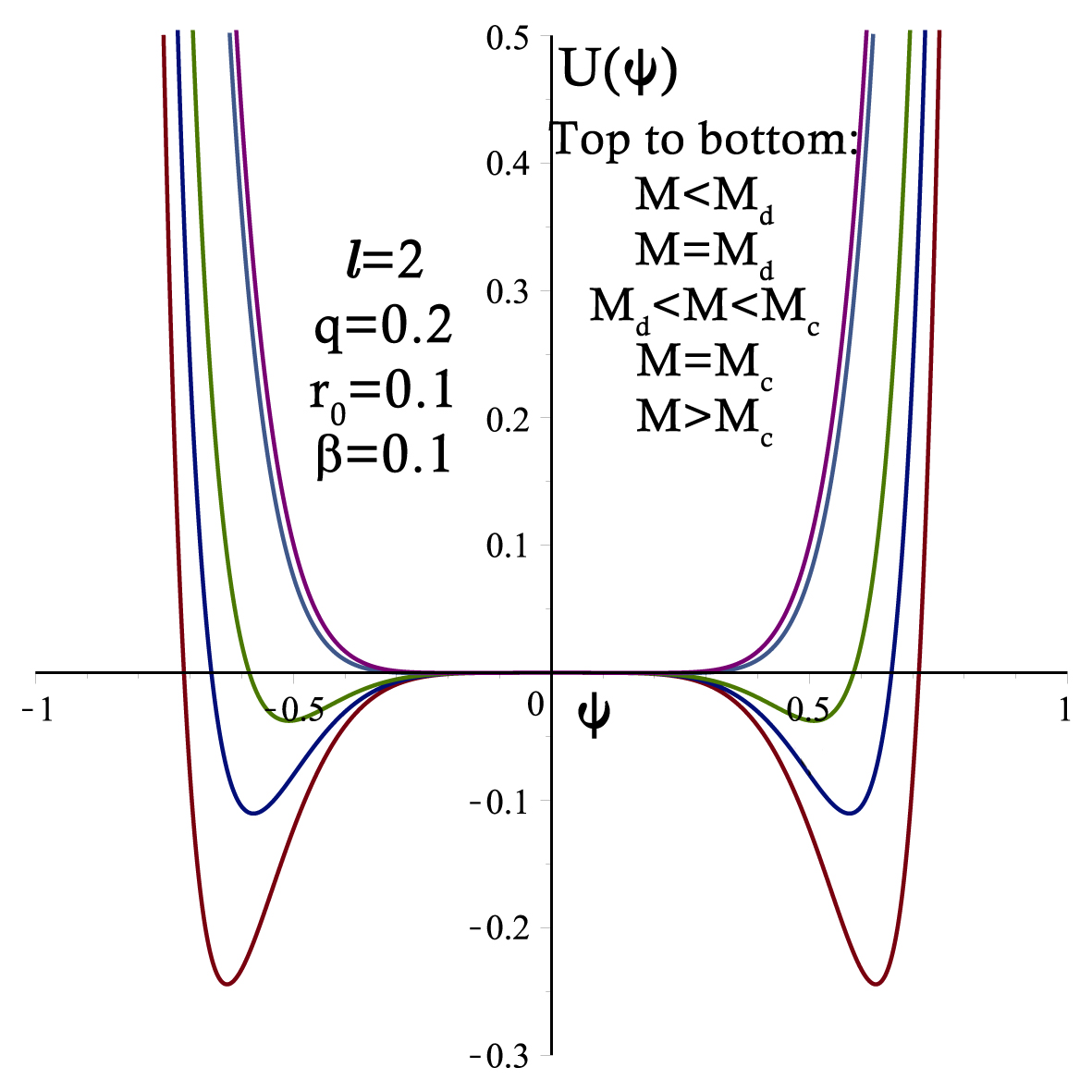}
\caption{The self-coupling potential $U\left( \protect\psi \right) $ versus $%
\protect\psi $ for different values of $M.$ We observe that the minima of
the potential occur when the $M_{d}<M$ which makes the central object a
black hole. For larger mass the minima of the potential are stronger. This
is valid for non-zero cosmological constant i.e., $\frac{1}{\ell ^{2}}\neq
0. $ }
\label{fig: 2}
\end{figure}
The general solution given in (12), depends on the free parameters and
non-zero cosmological constant. It admits one single horizon if $M_{c}\leq
M, $ two horizons if $M_{d}<M<M_{c},$ one degenerate horizon if $M=M_{d}$
and no horizon if $M<M_{d}$. We comment that, $M_{c}$ is found analytically
and is expressed as%
\begin{equation}
M_{c}=2q^{2}\ln \left( 1+\frac{2}{q\beta }\right) +\frac{q^{2}}{1+\beta ^{2}}
\end{equation}%
while $M_{d}$ should be found numerically for each set of parameters. In
Fig. 1 and 2 we plot the metric function $f\left( r\right) $ versus $r$ and
the self-coupling potential $U\left( \psi \right) $ versus $\psi $ to show
the effect of mass in forming different cases. We observe that to have a
black hole we must have a minimum mass and to have two absolute minimum
points for the potential the solution must be a black hole which means that $%
M>M_{d}.$ For the case in which the event horizon is present, the Hawking
temperature may be determined in terms of the radius of the event horizon $%
r_{h}.$ The explicit form of it is expressed as%
\begin{multline}
T_{H}=\frac{f^{\prime }\left( r_{h}\right) }{4\pi }= \\
\frac{r_{h}\left( q\beta +\frac{r_{h}^{2}}{\left( q\beta +\eta \right) }%
\right) \left( r_{0}+r_{h}\right) }{\eta \left( 2r_{0}+3r_{h}\right) \pi }%
\left( \frac{r_{0}\ln \left( \frac{q\beta +T}{r_{h}}\right) }{\beta ^{3}q}+%
\frac{3}{2\ell ^{2}}\right) \\
-\frac{q\left( r_{0}+3r_{h}\right) \left( r_{0}+r_{h}\right) }{\beta \left(
2r_{0}+3r_{h}\right) \pi r_{h}} \\
+\frac{\left[ 3r_{h}^{2}\left( r_{h}-\eta \right) +\left( r_{0}-3q\beta
\right) r^{2}\right] \left( r_{0}+r_{h}\right) }{\beta ^{2}\left(
2r_{0}+3r_{h}\right) \pi r_{h}\left( q\beta +\eta \right) },
\end{multline}%
in which $\eta =\sqrt{q^{2}\beta ^{2}+r_{h}^{2}}$. Fig. 3 displays the
effect of $r_{0}$ in $T_{H}.$ 
\begin{figure}[h]
\includegraphics[width=65mm,scale=0.7]{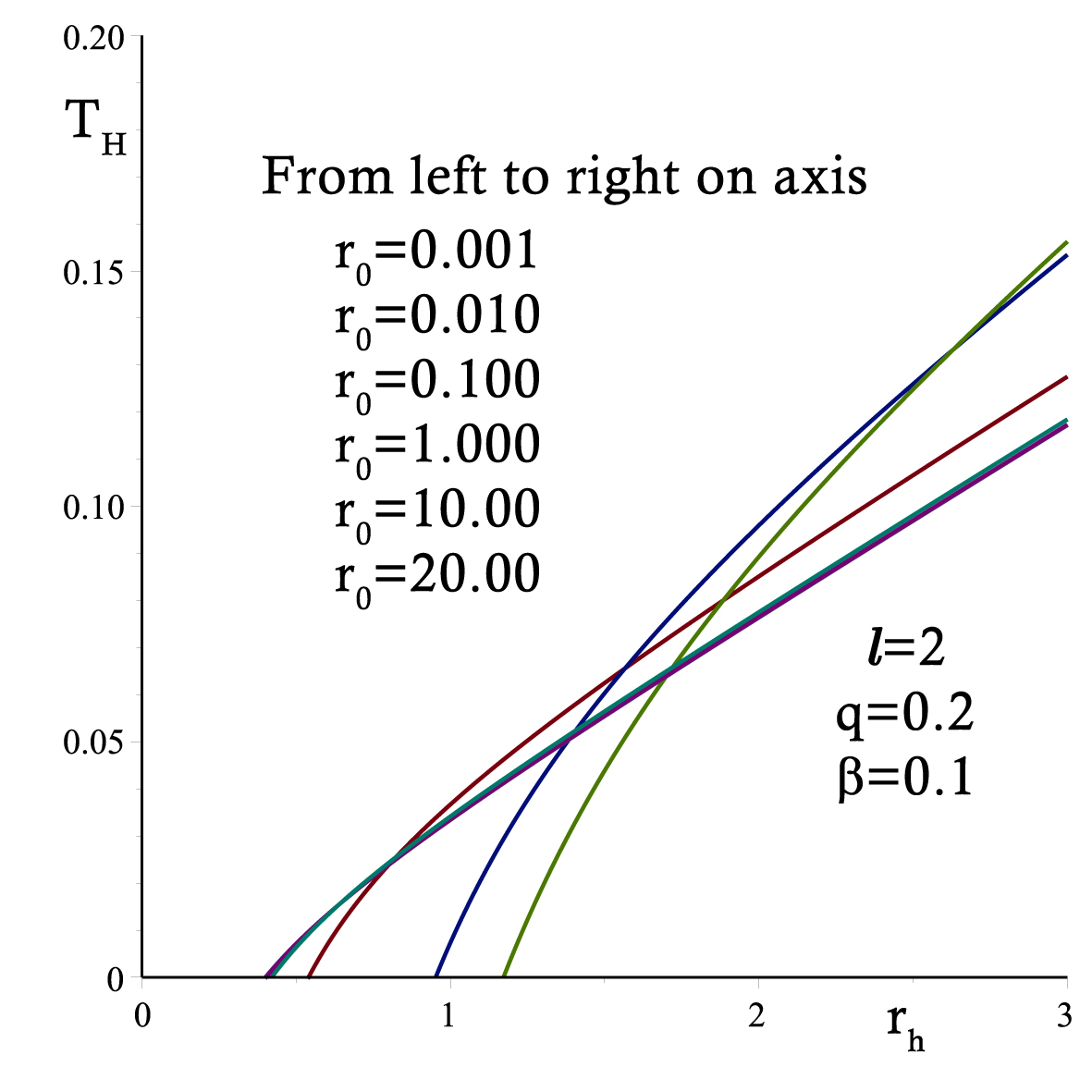}
\caption{Hawking temperature $T_{H}$ in terms of the radius of event horizon 
$r_{h}$ for different values of $r_{0}.$ We note that $r_{0}$ represents the
scalar field $\protect\psi $ and as $r_{0}$ goes to zero the solution
reduces to Einstein-Born-Infeld black hole studied by Cataldo et. al in 
\protect\cite{Cataldo}. Let us note that the negative temperature
corresponds to the non-black hole solution and therefore they are excluded.}
\label{fig: 3}
\end{figure}

\section{A dynamic solution}

The solution found in previous chapter for the metric function i.e., Eq.
(12) consists of five parameters which are $\beta ,$ $r_{0},$ $M,$ $q$ and $%
\ell ^{2}.$ These parameters also appear in the potential $V\left( \psi
\right) $ which makes our solution a rigid solution i.e., when $V\left( \psi
\right) $ is introduced in the action there is only one unique solution for
the metric function whose free parameters have already been chosen.
Therefore finding a dynamic solution which admits at least one parameter
free is needed. This task can be done by a redefinition of the free
parameters in the form of $V\left( \psi \right) $ given in Eq. (13).
Accordingly we introduce%
\begin{equation}
M=r_{0}^{2}\left( m-2Q^{2}\ln \left( \frac{r_{0}}{2+r_{0}Q\beta }\right)
\right)
\end{equation}%
and%
\begin{equation}
q=r_{0}Q
\end{equation}%
in which $m$ and $Q$ are two new parameters. Upon these change of
parameters, the potential and the metric function become%
\begin{multline}
V\left( \psi \right) = \\
\left( \frac{2Q^{2}}{3}\left[ \frac{1}{2\left( 1+\beta ^{2}\right) }-\ln
\left( \frac{1-\psi ^{2}+\sqrt{\mathcal{K}}}{\psi ^{2}}\right) \right] +%
\frac{1}{\ell ^{2}}-\frac{m}{3}\right) \psi ^{6}+ \\
\frac{2\left( 2Q^{2}\beta ^{2}+3Q\beta +2\right) \psi ^{8}+2\left( \beta
^{2}Q^{2}\psi ^{4}-3\psi ^{4}+1\right) \psi ^{2}}{3\beta ^{2}\left( \psi
^{2}Q\beta +\sqrt{\mathcal{K}}\right) }+ \\
\frac{2\left( 2\beta Q\psi ^{2}+3\psi ^{2}+Q\beta \right) \psi ^{4}\sqrt{%
\mathcal{K}}}{3\beta ^{2}\left( \psi ^{2}Q\beta +\sqrt{\mathcal{K}}\right) }-%
\frac{1}{\ell ^{2}}+ \\
\frac{2\left( \psi ^{6}-1\right) }{3Q\beta ^{3}}\ln \left( \frac{\psi
^{2}Q\beta +\sqrt{\mathcal{K}}}{1-\psi ^{2}}\right)
\end{multline}%
and%
\begin{multline}
f\left( r\right) =-\frac{r_{0}^{2}\left( 2r_{0}+3r\right) }{3r}\left[ m-%
\frac{Q^{2}}{1+\beta ^{2}}+2Q^{2}\ln \left( \frac{r+\sqrt{\mathcal{H}}}{r_{0}%
}\right) \right] + \\
r^{2}\left[ \frac{2}{3\beta ^{3}Q}\ln \left( \frac{r_{0}Q\beta +\sqrt{%
\mathcal{H}}}{r}\right) +\frac{1}{\ell ^{2}}+\frac{2}{\beta ^{2}}\right] -%
\frac{2\left( r_{0}+3r\right) }{3\beta ^{2}}\sqrt{\mathcal{H}},
\end{multline}%
in which $\mathcal{K}=$ $\left( 1-\psi ^{2}\right) ^{2}+\beta ^{2}Q^{2}\psi
^{4}$ and $\mathcal{H}=r^{2}+r_{0}^{2}Q^{2}\beta ^{2}.$ The form of the
scalar field remains the same as it is given in (11). We see that the
potential $V\left( \psi \right) $ is independent of $r_{0}$ and it consists
of only four parameters which are $\beta ,$ $m,Q$ and $\ell ^{2}$ while in
the metric function there are five parameters including $r_{0}$, $\beta ,$ $%
m,Q$ and $\ell ^{2}.$ In this sense we have a dynamic solution with respect
to $r_{0}.$ 

\begin{figure}[h]
\includegraphics[width=65mm,scale=0.7]{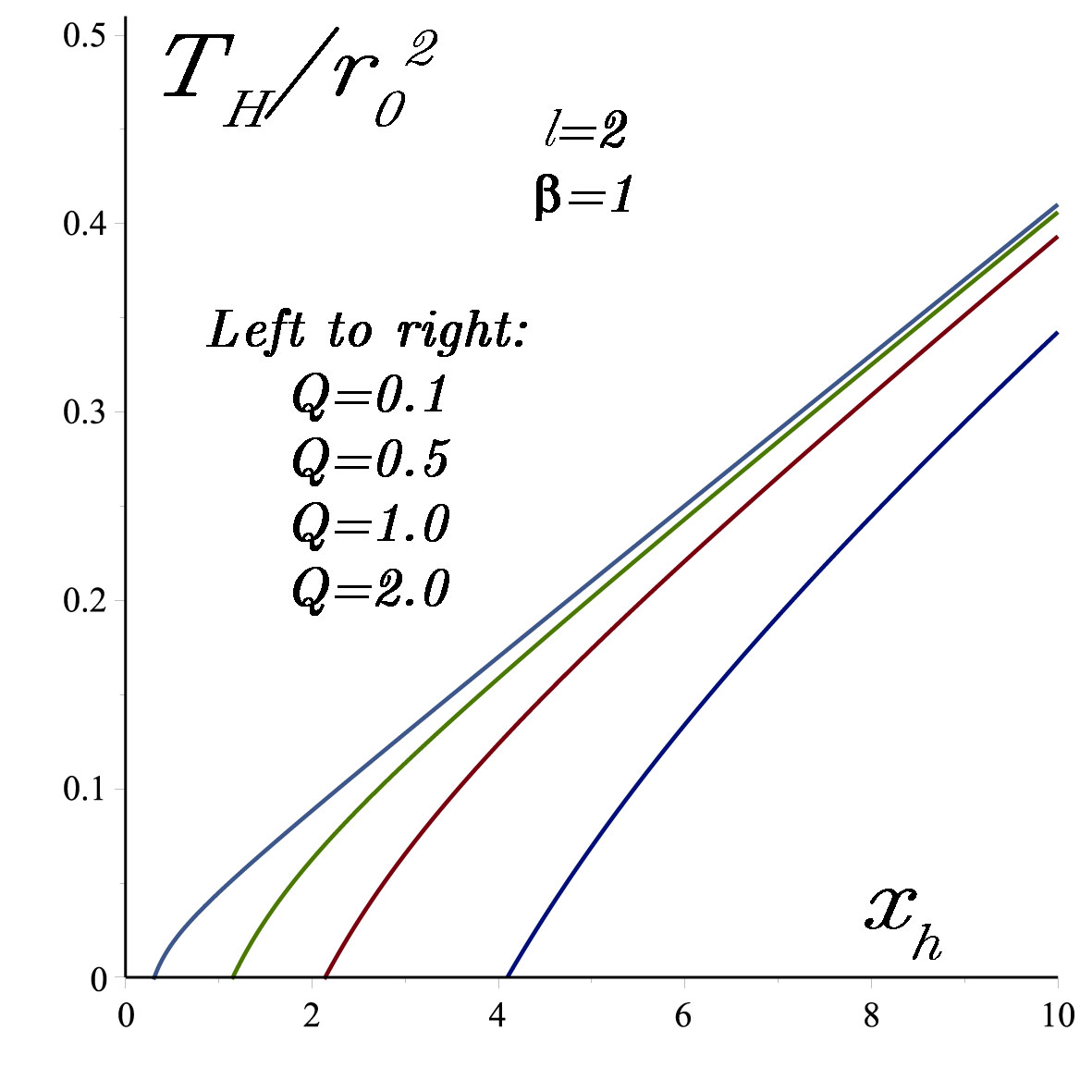}
\caption{Hawking temperature $\frac{T_{H}}{r_{0}^{2}}$ in terms of the
radius of event horizon $x_{h}=\frac{r_{h}}{r_{0}}$ for varies values of $Q$
and fixed values of $\ell ^{2}$ and $\protect\beta .$ It is remarkable to
observe that the only free parameter in the metric function, i.e., $r_{0}$
appears as a scale parameter. }
\label{fig: 4}
\end{figure}
\begin{figure}[h]
\includegraphics[width=65mm,scale=0.7]{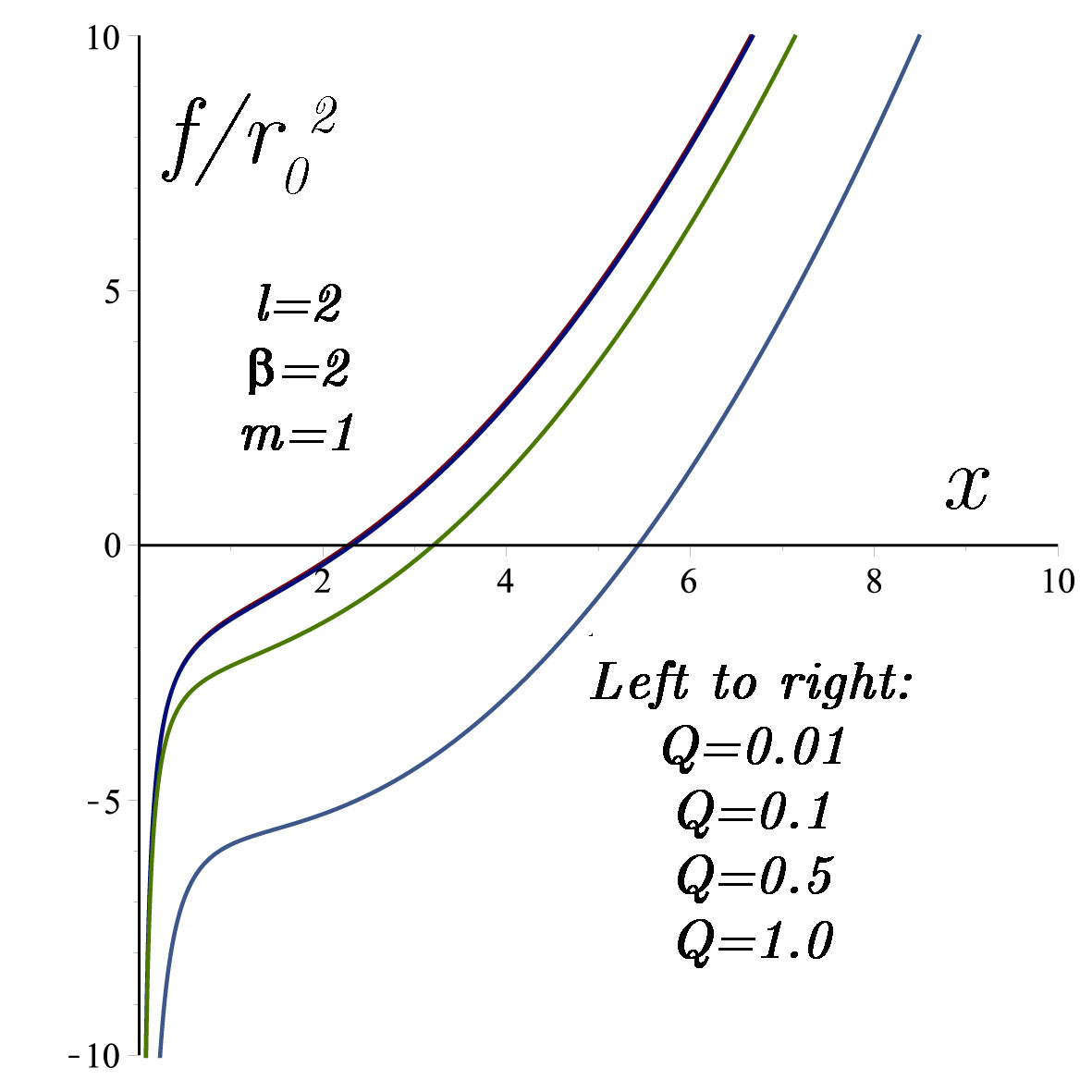}
\caption{Metric function $\frac{f}{r_{0}^{2}}$ in terms of $x=\frac{r}{r_{0}}
$ for fixed values of $m,\protect\beta ,\ell $ and various values of $Q.$ We
see that the scalar parameter $r_{0},$ acts as a scale parameter. Also we
see that the metric presents a black hole.}
\label{fig: 5}
\end{figure}

\begin{figure}[h]
\includegraphics[width=65mm,scale=0.7]{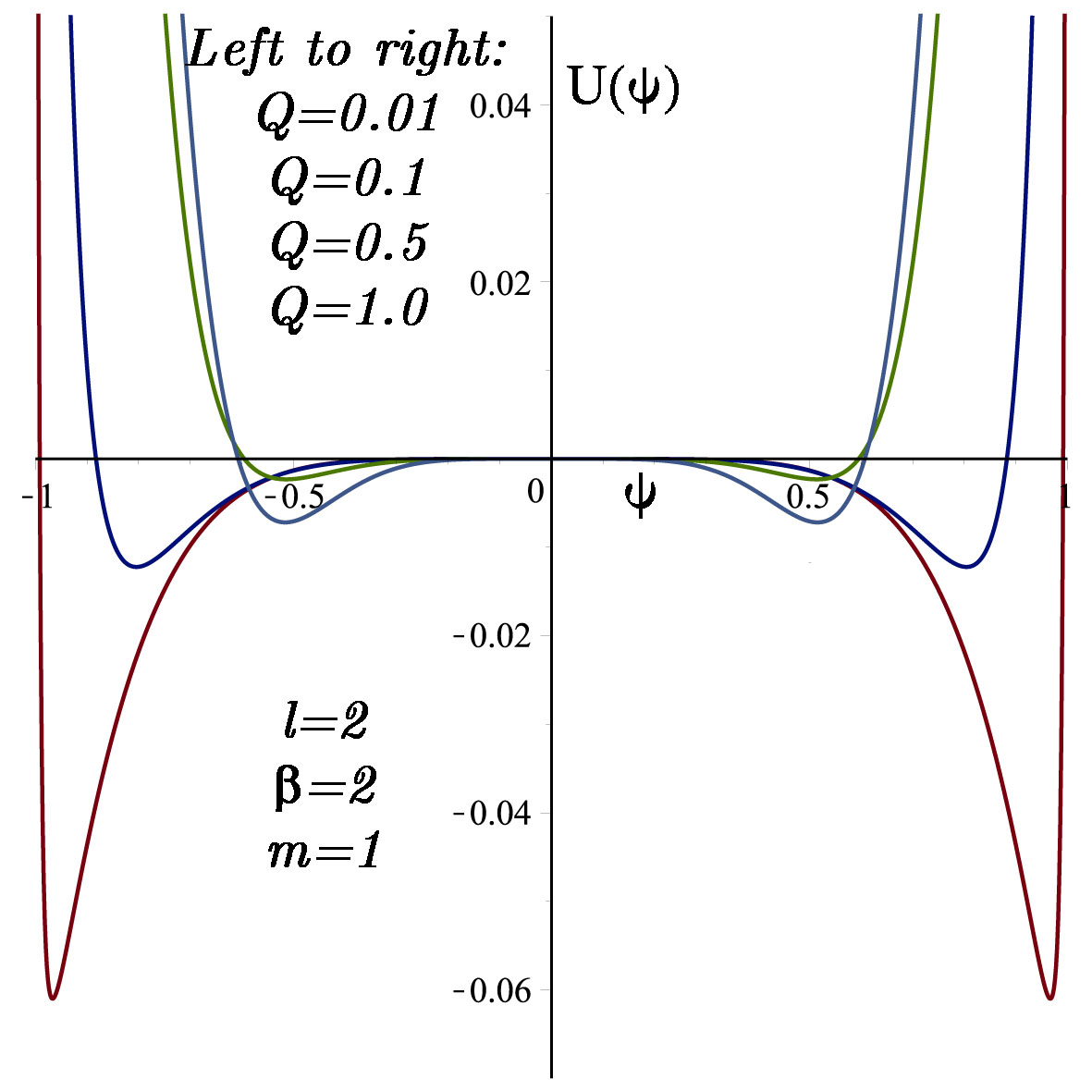}
\caption{$U\left( \protect\psi \right) $ in terms of $\protect\psi $ for
various values of $Q.$ The other parameters are fixed as shown in the
figure. For $Q$ very small, $U$ is proportional to $\protect\psi ^{6}$ while
for larger $Q$ deviation from $\protect\psi ^{6}$ occurs.}
\end{figure}
Let us add that the dynamic potential found above in the limit $Q\rightarrow
0$ admits%
\begin{equation}
\lim_{Q\rightarrow 0}V\left( \psi \right) =\frac{1}{\ell ^{2}}+\left( \frac{1%
}{\ell ^{2}}-\frac{\beta ^{2}m-12}{3\beta ^{2}}\right) \psi ^{6}
\end{equation}%
and in order to get Einstein-scalar solution one must consider the limit
when $\beta \rightarrow \infty .$ The result turns out to be 
\begin{equation*}
\lim_{\substack{ Q\rightarrow 0 \\ \beta \rightarrow \infty }}V\left( \psi
\right) =\frac{1}{\ell ^{2}}+\left( \frac{1}{\ell ^{2}}-\frac{m}{3}\right)
\psi ^{6}
\end{equation*}%
which is comparable with the potential studied in \cite{39} provided $\left( 
\frac{1}{\ell ^{2}}-\frac{m}{3}\right) \psi ^{6}\equiv \alpha _{3}\phi ^{6},$
or equivalently $\alpha _{3}=\frac{1}{512}\left( \frac{1}{\ell ^{2}}-\frac{m%
}{3}\right) $. We note that due to rescaling of scalar field used in our
calculation and the one used in \cite{39} the two are proportional i.e., $%
\psi ^{2}=\frac{1}{8}\phi ^{2}.$ In addition, the metric function in the
same limit admits 
\begin{equation}
\lim_{\substack{ Q\rightarrow 0 \\ \beta \rightarrow \infty }}f\left(
r\right) =\frac{r^{2}}{\ell ^{2}}-\frac{r_{0}^{2}m}{3}\left( 3+\frac{2r_{0}}{%
r}\right) 
\end{equation}%
which is in perfect match with the metric function Eq. (36) found in \cite%
{39}.

The solution (38) is a black hole solution whose Hawking temperature is
obtained as%
\begin{multline}
\frac{T_{H}}{r_{0}^{2}}=\frac{\left( 1+x_{h}\right) x_{h}\left(
1+3x_{h}\right) }{Q\beta ^{3}\pi \left( 2+3x_{h}\right) \left( 1+\sqrt{1+%
\frac{x_{h}^{2}}{Q^{2}\beta ^{2}}}\right) }+ \\
\frac{x_{h}\left( 1+x_{h}\right) }{Q\beta ^{3}\pi \left( 2+3x_{h}\right) }%
\ln \left( \frac{Q\beta }{x_{h}}\left( 1+\sqrt{1+\frac{x_{h}^{2}}{Q^{2}\beta
^{2}}}\right) \right) + \\
\frac{3Q\left( 1+x_{h}\right) \left( \frac{x_{h}^{2}}{Q\beta }-x_{h}-\frac{1%
}{3}\right) }{\pi x\beta \left( 2+3x_{h}\right) }+\frac{3}{2}\frac{\left(
1+x_{h}\right) x_{h}}{\pi \left( 2+3x_{h}\right) \ell ^{2}},
\end{multline}%
in which $x_{h}=\frac{r_{h}}{r_{0}}.$ Fig. 4 is a plot of $\frac{T_{H}}{%
r_{0}^{2}}$ versus $x_{h}$ for different values of $Q$ but fixed values for
other free parameters. As one observes from the latter expression, the free
parameter of the metric function i.e., $r_{0}$ acts as a universal scaling.
This is not surprising because the metric function may also be written as%
\begin{multline}
\frac{f}{r_{0}^{2}}=-\frac{2Q^{2}\left( \frac{2}{3}+x\right) \ln \left( x+%
\sqrt{Q^{2}\beta ^{2}+x^{2}}\right) }{x}+ \\
\frac{2x^{2}\ln \left( \frac{Q\beta +\sqrt{Q^{2}\beta ^{2}+x^{2}}}{x}\right) 
}{3\beta ^{3}Q}+x^{2}\left( \frac{1}{\ell ^{2}}+\frac{2}{\beta ^{2}}\right) -
\\
\frac{\left( 2+3x\right) }{3x}\left( m-\frac{Q^{2}}{1+\beta ^{2}}\right) -%
\frac{2\left( 1+3x\right) \sqrt{Q^{2}\beta ^{2}+x^{2}}}{3\beta ^{2}}
\end{multline}%
in which $x=\frac{r}{r_{0}}.$ In Fig. 5 we plot $\frac{f}{r_{0}^{2}}$ in
terms of $x$ for $m=1,$ $\beta =2$, $\ell =2$ and various values of $Q.$ We
also plot $U\left( \psi \right) $ versus $\psi $ for various parameters in
Fig. 6. In this figure the general behavior of the potential is depicted in
terms of $\psi $ but different values for $Q.$ For small $Q$ it is
proportional to $\psi ^{6}$ which is in agreement with Ref. \cite{39}.

\section{A general formalism}

In $3+1-$dimensions the first scalar-tensor black holes were studied in \cite%
{40,41,42}. Following the method employed in \cite{40,41,42} we give a
general picture of the scalar-tensor black holes in $2+1-$dimensions without
solving the field equations explicitly. We start with the $2+1-$dimensional
scalar-tensor action given by%
\begin{multline}
S=\frac{1}{16\pi G}\int d^{3}x\sqrt{-\tilde{g}}\times  \\
\left[ W\left( \Phi \right) \tilde{R}-Z\left( \Phi \right) \tilde{g}^{\mu
\nu }\partial _{\mu }\Phi \partial _{\nu }\Phi -2U\left( \Phi \right) \right]
+S_{m}\left[ \Psi _{m};\tilde{g}_{\mu \nu }\right] 
\end{multline}%
in which $\Phi $ is the scalar field, $W\left( \Phi \right) ,$ $Z\left( \Phi
\right) $ and $U\left( \Phi \right) $ are functions of $\Phi $ and 
\begin{equation}
S_{m}\left[ \Psi _{m};\tilde{g}_{\mu \nu }\right] =\frac{1}{16\pi G}\int
d^{3}x\sqrt{-\tilde{g}}L\left( F\right) 
\end{equation}%
is the matter field ($\Psi _{m}$)-coupled NED action. Then we apply the
following conformal transformation \cite{43}%
\begin{equation}
g_{\mu \nu }=W\left( \Phi \right) ^{2}\tilde{g}_{\mu \nu }
\end{equation}%
in order to go from the Jordan frame with metric tensor $\tilde{g}_{\mu \nu }
$ to Einstein frame with metric tensor $g_{\mu \nu }.$ We introduce a
dilaton field $\varphi $ satisfying%
\begin{equation}
\left( \frac{d\ln W\left( \Phi \right) }{d\Phi }\right) ^{2}+\frac{Z\left(
\Phi \right) }{2W}=\left( \frac{d\varphi }{d\Phi }\right) ^{2},
\end{equation}%
with the new potential%
\begin{equation}
V\left( \varphi \right) =\frac{U\left( \Phi \right) }{2W\left( \Phi \right)
^{3}}
\end{equation}%
and with the new notation 
\begin{equation}
A\left( \varphi \right) =\frac{1}{W\left( \Phi \right) }
\end{equation}%
the action (43) in Einstein frame takes the form%
\begin{multline}
S=\frac{1}{16\pi G}\int d^{3}x\sqrt{-g}\times  \\
\left[ R-2g^{\mu \nu }\partial _{\mu }\varphi \partial _{\nu }\varphi
-4V\left( \varphi \right) +A\left( \varphi \right) ^{3}L\left( X\right) %
\right] .
\end{multline}%
We note that in order that the dilaton carries positive energy in $2+1-$%
dimensions we must have both conditions $W\left( \Phi \right) >0$ and
consequently from (46) 
\begin{equation}
2\left( \frac{dW}{d\Phi }\right) ^{2}+ZW\geq 0
\end{equation}%
satisfied. Comparing the actions (43) and (1), one reads from (1), $W=1-\Phi
^{2}$ and $Z=8$ with $\Phi ^{2}$ given in (11). One can easily check that
both conditions given above are satisfied (note that in (1) $\psi $ plays
the role of $\Phi $ in (43)). Let's add also that the matter action gets the
form 
\begin{equation}
S_{m}=\frac{1}{16\pi G}\int d^{3}xA\left( \varphi \right) ^{3}\sqrt{-g}%
L\left( X\right) 
\end{equation}%
in which%
\begin{equation}
X=A\left( \varphi \right) ^{-4}F_{\mu \nu }F_{\alpha \beta }g^{\alpha \mu
}g^{\beta \nu }.
\end{equation}%
Using the action (49) and (51) together, one finds the field equations given
by%
\begin{multline}
R_{\mu \nu }=2\partial _{\mu }\varphi \partial _{\nu }\varphi +4V\left(
\varphi \right) g_{\mu \nu }- \\
\frac{2L_{X}}{A}\left( F_{\mu \beta }F_{\nu }^{\beta }-g_{\mu \nu }F_{\alpha
\beta }F^{\alpha \beta }\right) -A\left( \varphi \right) ^{3}L\left(
X\right) g_{\mu \nu },
\end{multline}%
\begin{equation}
d\left( \frac{\mathbf{\tilde{F}}L_{X}}{A\left( \varphi \right) }\right) =0
\end{equation}%
and%
\begin{equation}
\square \varphi =\frac{dV\left( \varphi \right) }{d\varphi }+\frac{d\ln
A\left( \varphi \right) }{d\varphi }A\left( \varphi \right) ^{3}\left(
XL_{X}-\frac{3}{4}L\right) 
\end{equation}%
in which $L_{X}=\frac{dL}{dX}.$ The static and spherically symmetric line
element is chosen to be%
\begin{equation}
ds^{2}=-f\left( r\right) e^{-2\delta \left( r\right) }dt^{2}+\frac{dr^{2}}{%
f\left( r\right) }+r^{2}d\theta ^{2}
\end{equation}%
in which $f$ and $\delta $ are only function of $r$. This line element with
two unknown functions $f\left( r\right) $ and $\delta \left( r\right) $ with
the field Eqs. (53)-(55) constitute our $2+1-$dimensional general
scalar-tensor field equations coupled to an NED Lagrangian. The field
equations can be written explicitly as%
\begin{equation}
\frac{d\delta }{dr}=-2r\left( \frac{d\varphi }{dr}\right) ^{2}
\end{equation}%
\begin{equation}
\frac{df}{dr}=-r\left( 4V+2f\left( \frac{d\varphi }{dr}\right)
^{2}+A^{3}\left( 2XL_{X}-L\right) \right) 
\end{equation}%
and%
\begin{multline}
\frac{d}{dr}\left( re^{-\delta }f\frac{d\varphi }{dr}\right) = \\
re^{-\delta }\left( \frac{dV}{d\varphi }+\frac{d\ln A}{d\varphi }A\left(
\varphi \right) ^{3}\left( XL_{X}-\frac{3}{4}L\right) \right) .
\end{multline}%
It is observed that for static, spherical symmetry with $\delta \left(
r\right) =cons.$, $\varphi $ reduces to a constant which can be considered
as the cosmological constant. For a general scalar-tensor model, however, $%
\delta \left( r\right) \neq cons.$ must be determined as well.

Let's add that the nonlinear Maxwell equation with an electric field ansatz 
\begin{equation}
\mathbf{F}=E\left( r\right) dt\wedge dr
\end{equation}%
implies%
\begin{equation}
E^{2}=\frac{C_{0}^{2}e^{-2\delta }A^{4}}{A^{2}+C_{0}^{2}\beta ^{2}}
\end{equation}%
in which $C_{0}^{2}$ is an integration constant. This also implies that $%
F_{\mu \nu }F^{\mu \nu }=\frac{-2C_{0}^{2}A^{4}}{A^{2}+C_{0}^{2}\beta ^{2}}$
and $X=\frac{-2C_{0}^{2}}{A^{2}+C_{0}^{2}\beta ^{2}}.$ With $V=0$ one finds
that the right hand side of (59) is positive if $\frac{dA}{d\varphi }>0.$
This is what we would like to consider and then $\frac{d}{dr}\left(
re^{-\delta }f\frac{d\varphi }{dr}\right) >0$ which means that if $\varphi $
is increasing / decreasing function with respect to $r$ the metric function $%
f$ admits at most a single root to be identified as the event horizon of the
black hole. Note that $XL_{X}-\frac{3}{4}L$ for $L$ given by (8) is positive
definite. Different ansatzes other than (60) leads naturally to new
solutions which is not our concern here. Once more we refer to \cite%
{40,41,42} for the details of the requirements we have applied here.

\section{Conclusion}

A field theory model of Einstein-Scalar-Born-Infeld is considered in $2+1-$%
dimensions. In the first part of the paper we obtain a rigid solution to the
field equations. Depending on the parameters this naturally admits black
hole and non-black hole solutions. This is depicted in Fig. 1 numerically,
in which the mass plays a crucial role. The scalar hair dependence of both
the self-interacting potential $U\left( \psi \right) $ and the Hawking
temperature are also displayed. The self-interacting potential $U\left( \psi
\right) $ is highly non-linear with reflection symmetry $U\left( \psi
\right) =U\left( -\psi \right) $. When $U\left( \psi \right) $ admits no
minima it asymptotes to an infinite potential well of quantum mechanics.
With the proper choice of parameters the minima are produced as displayed in
Fig. 2. In the limit of (BI parameter) $\beta \rightarrow 0$ our results
reduce mainly, up to minor scalings, to the ones obtained in Ref. \cite{8}.
Our contribution therefore is to extend the hairy black holes of
linear-Maxwell theory to nonlinear BI theory in the presence of a
self-interacting scalar field. We must admit that exact solutions were
obtained at the price of tuning the integration constants. Without such
choices finding solution for such a non-linear model field theory remains
out of our reach. However, by using this solution a more physical, dynamic
solution can be constructed. Let's add that with the BI addition it is
observed from Eq. (32) that the singularity at $r=0$ modifies from $\frac{1}{%
r^{3}}$ of linear Maxwell theory \cite{8} to the form given by (32).

In the second part of the paper we use the rigid solution, found in the
first part to construct a dynamic black hole solution (38). The free
parameter in the dynamic metric function is the scalar charge $r_{0}.$ In
Figs. 4 and 5 we show that this parameter can be considered as a scaling
parameter. The form of the potential $U\left( \psi \right) $ is also
investigated in Fig. 6 and it is shown that the specific potential of $%
U\left( \psi \right) \sim \psi ^{6}$ studied in \cite{39} is also recovered.
A general discussion for scalar-tensor coupled NED theory is also included
in the paper briefly, leaving the details to a future correspondence.

\appendix*

\section{A}

The field equations explicitly become%
\begin{equation}
G_{t}^{t}-\tau _{t}^{t}-T_{t}^{t}+V=0,
\end{equation}%
\begin{equation}
G_{r}^{r}-\tau _{r}^{r}-T_{r}^{r}+V=0,
\end{equation}%
\begin{equation}
G_{\theta }^{\theta }-\tau _{\theta }^{\theta }-T_{\theta }^{\theta }+V=0
\end{equation}%
and%
\begin{equation}
\frac{1}{r}\left[ f\psi ^{\prime }+rf^{\prime }\psi ^{\prime }+rf\psi
^{\prime \prime }\right] -\frac{1}{8}R\psi -\frac{1}{8}\frac{dV}{d\psi }=0.
\end{equation}%
Herein 
\begin{equation}
G_{t}^{t}=G_{r}^{r}=\frac{f^{\prime }}{2r},\text{ }G_{\theta }^{\theta }=%
\frac{1}{2}f^{\prime \prime },
\end{equation}%
\begin{equation}
\tau _{t}^{t}=\frac{1}{2r}\left( -4rf\psi ^{\prime 2}+4rf\psi \psi ^{\prime
\prime }+2\psi \psi ^{\prime }\left( 2f+rf^{\prime }\right) +f^{\prime }\psi
^{2}\right)
\end{equation}%
\begin{equation}
\tau _{r}^{r}=\frac{\left( f^{\prime }\psi +4f\psi ^{\prime }\right) \left(
2r\psi ^{\prime }+\psi \right) }{2r}
\end{equation}%
\begin{equation}
\tau _{\theta }^{\theta }=-2f\psi ^{\prime 2}+2f^{\prime }\psi \psi ^{\prime
}+2f\psi \psi ^{\prime \prime }+\frac{1}{2}f^{\prime \prime }\psi ^{2}
\end{equation}%
\begin{equation}
R=-\frac{rf^{\prime \prime }+2f^{\prime }}{r}
\end{equation}%
\begin{equation}
T_{t}^{t}=T_{r}^{r}=\frac{2\beta \left( -r^{2}\beta ^{2}+\beta r\sqrt{%
q^{2}+r^{2}\beta ^{2}}-q^{2}\right) }{r\sqrt{q^{2}+r^{2}\beta ^{2}}}
\end{equation}%
and%
\begin{equation}
T_{\theta }^{\theta }=\frac{2\beta ^{2}\left( -\beta r+\sqrt{%
q^{2}+r^{2}\beta ^{2}}\right) }{\sqrt{q^{2}+r^{2}\beta ^{2}}}.
\end{equation}%
After some simplification the field equation can be written as%
\begin{multline}
\frac{1}{2r}\left( -4rf\psi \psi ^{\prime \prime }+4rf\psi ^{\prime 2}-2\psi
\psi ^{\prime }\left( 2f+rf^{\prime }\right) -f^{\prime }\psi ^{2}\right) \\
+\frac{f^{\prime }}{2r}-T_{t}^{t}+V=0
\end{multline}%
\begin{equation}
\frac{f^{\prime }}{2r}-\frac{\left( 2r\psi ^{\prime }+\psi \right) \left(
4f\psi ^{\prime }+f^{\prime }\psi \right) }{2r}-T_{t}^{t}+V=0
\end{equation}%
\begin{multline}
\frac{1}{2}f^{\prime \prime }+2f\psi ^{\prime 2}-2f^{\prime }\psi \psi
^{\prime }-2f\psi \psi ^{\prime \prime }- \\
\frac{1}{2}f^{\prime \prime }\psi ^{2}-T_{\theta }^{\theta }+V=0
\end{multline}%
and Eq. (A.4). Next, we subtract (A.13) from (A.12) which simply gives%
\begin{equation}
-\psi \psi ^{\prime \prime }+3\psi ^{\prime 2}=0.
\end{equation}%
This equation admits a solution of the form 
\begin{equation}
\psi ^{2}=\frac{1}{c_{1}r+c_{2}}
\end{equation}%
in which $c_{1}$ and $c_{2}$ are two integration constants. Hence by
redefinition of constants one may write%
\begin{equation}
\psi ^{2}=\frac{\mu ^{2}}{1+\frac{r}{r_{0}}},
\end{equation}%
in which $\mu $ and $r_{0}>0$ are two new constants related to $c_{1}$ and $%
c_{2},$ both nonzero. Upon finding $\psi ^{2},$ one may subtract (A.12) from
(A.14) to find a differential equation for only $f\left( r\right) $ i.e.%
\begin{multline}
r\left( 1-\psi ^{2}\right) f^{\prime \prime }+\left[ \psi ^{2}-2r\psi \psi
^{\prime }-1\right] f^{\prime }+ \\
4f\psi \psi ^{\prime }+2r\left( T_{t}^{t}-T_{2}^{2}\right) =0,
\end{multline}%
or explicitly%
\begin{multline}
\frac{1}{2}\left( r+r_{0}\right) \left( r-r_{0}\left[ \mu ^{2}-1\right]
\right) rf^{\prime \prime }+ \\
\left( r_{0}\left( \mu ^{2}-1\right) \left( r+\frac{1}{2}r_{0}\right) -\frac{%
r^{2}}{2}\right) f^{\prime }- \\
r_{0}\mu ^{2}f+r\left( r+r_{0}\right) ^{2}\left( T_{t}^{t}-T_{2}^{2}\right)
=0.
\end{multline}%
In this DE there exist four parameters, $\beta ,$ $q,$ $\mu $ and $r_{0}.$%
The complete solution to this equation is complicated in general but by
setting $\mu =1$ a special solution interesting enough is given by Eq. (12)
in the text which includes two new integration constants that are shown by $%
\ell ^{2}$ and $M.$ Up to here, without specifying the form of the potential 
$V\left( \psi \right) $ we found $\psi $ and $f.$ Finally one may use one of
the Eqs. (A.12)-(A.14) to find the exact form of the potential $V\left( \psi
\right) $. The consistency of the metric function, scalar field and
potential can be seen when they satisfy perfectly the last equation (A.4).

\bigskip

\end{document}